\documentclass[conference]{IEEEtran}
\usepackage{cite}

\usepackage{graphicx,color,epsfig,rotating,subfigure}
\usepackage{amsfonts,amsmath,amssymb}
\usepackage{algorithm}
\usepackage{subfigure}
\usepackage{algpseudocode}

\long\def\comment#1{}

\newfont{\bbb}{msbm10 scaled 700}

\newfont{\bb}{msbm10 scaled 1100}

\newcommand{\uv}{{\bf u}}
\newcommand{\wv}{{\bf w}}

\newcommand{\xv}{{\bf x}}
\newcommand{\yv}{{\bf y}}

\newcommand{\Bc}{{\cal B}}

\newcommand{\Ic}{{\cal I}}

\newcommand{\Tc}{{\cal T}}

\newcommand{\Vc}{{\cal V}}

\newcommand{\SNR}{{\sf SNR}}
\newcommand{\INR}{{\sf INR}}

\newcommand{\eqdef}{\stackrel{\Delta}{=}}

\newtheorem{definition}{Definition}
\newtheorem{theorem}{Theorem}

\newtheorem{corollary}{Corollary}

\newtheorem{remark}{Remark}

\begin{document}

\title{On the Achievable Rates of Multihop Virtual Full-Duplex Relay Channels}
\author{\authorblockN{Song-Nam Hong\authorrefmark{1}, Ivana Mari\'c\authorrefmark{1}, Dennis Hui\authorrefmark{1} and Giuseppe Caire\authorrefmark{2}}
\authorblockA{\authorrefmark{1} Ericsson Research, San Jose, CA, \\{\it email:(songnam.hong, ivana.maric, dennis.hui)@ericsson.com}}
\authorblockA{\authorrefmark{2} Technical University of Berlin, Germany, \\{\it email:caire@tu-berlin.de}}
}
\maketitle

\begin{abstract} We study a multihop ``virtual" full-duplex relay channel as a special case of a general multiple multicast relay network. For such channel, quantize-map-and-forward (QMF) (or noisy network coding (NNC)) achieves the cut-set upper bound within a constant gap where the gap grows {\em linearly} with the number of relay stages $K$. However, this gap may not be negligible for the systems with multihop transmissions (i.e., a wireless backhaul operating at higher frequencies). We have recently attained an improved result to the capacity scaling where the gap grows {\em logarithmically} as $\log{K}$, by using an optimal quantization at relays and by exploiting relays' messages (decoded in the previous time slot) as side-information. In this paper, we further improve the performance of this network by presenting a {\em mixed} scheme where each relay can perform either decode-and-forward (DF) or QMF with possibly rate-splitting. We derive an achievable rate and show that the proposed scheme outperforms the optimized QMF. Furthermore, we demonstrate that this performance improvement increases with $K$.

\end{abstract}

\begin{IEEEkeywords}
Multihop relay networks, wireless backhaul, quantize-map-and-forward
\end{IEEEkeywords}

\section{Introduction}\label{sec:Intro}

Recent works have demonstrated the practical feasibility of full-duplex relays through the suppression of self-interference in a mixed analog-digital fashion in order to avoid the problem of receiver power saturation \cite{Duarte,Choi}. These architectures are based on some form of analog self-interference cancellation, followed by digital self-interference cancellation in the baseband domain. In some of these architectures, the self-interference cancellation in the analog domain is obtained by transmitting with multiple antennas such that the signals transmitted over different antennas superimpose in opposite phases and therefore cancel each other at the receiving antennas. Building on the idea of using multiple antennas to cope with the isolation of the receiver from the transmitter, we may consider a ``distributed version" of such approach where the transmit and receive antennas belong to physically separated nodes. This has the advantage that each of such nodes operates in conventional half-duplex mode. Furthermore, by allowing a large physical separation between nodes, the problem of receiver power saturation is eliminated.

\begin{figure}
\centerline{\includegraphics[width=8cm]{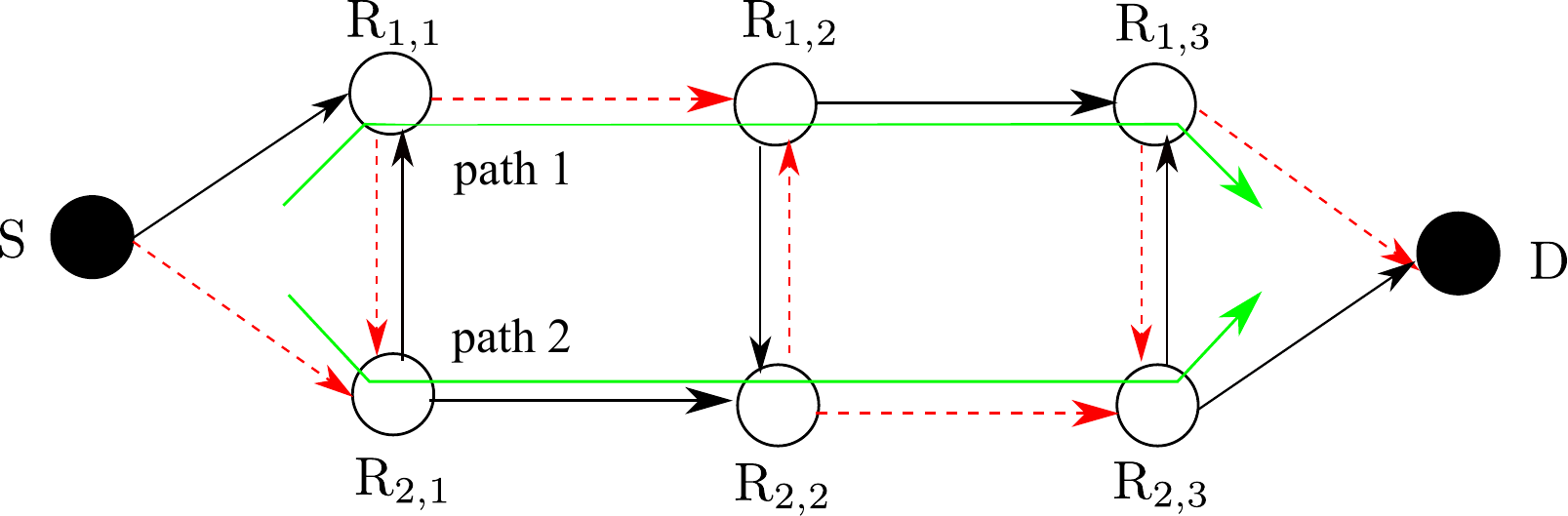}}
\caption{Multihop virtual full-duplex relay channels when $K=3$ (i.e., 4-hop relay network). Black-solid lines are active for every odd time slot and
red-dashed lines are active for every even time slot.}
\label{K-Hops}
\end{figure}

Motivated by the distributed approach, we introduce a communication scheme that utilizes ``virtual" full-duplex relays, each consisting of two half-duplex relays. In this configuration, each relay stage is formed of at least two half-duplex relays, used alternatively in transmit and receive modes, such that while one relay transmits its signal to the next stage, the other relay receives a signal from the previous stage. The role of the relays is swapped at the end of each time interval (see Fig.~\ref{K-Hops}). This relaying operation is known as ``successive relaying" \cite{Rezaei}. In this way, the source can send a new message to the destination at every time slot as if full-duplex relays are used. Every two consecutive source messages will travel via two alternate  disjoint paths of relays. In \cite{Bagheri}, 2-hop model has been studied, showing that dirty paper coding (DPC) achieves the performance of {\em ideal} full-duplex relay since the source can completely eliminate the ``known" interference at intended receiver. However, DPC is no longer applicable in a multihop network model shown in Fig.~\ref{K-Hops} since a transmit relay has no knowledge on interference signals at other stages \cite{Muthuramalingam}. Thus, finding an optimal strategy for the multihop models is still an open problem.

Since the multihop model is a special case of a single-source single-destination (non-layered) network, QMF \cite{Avestimehr} (or  NNC  \cite{Lim, Hou}) can be applied to this model. By setting the quantization distortion levels to be at the background noise level, these schemes achieve the capacity within a constant gap that scales {\it linearly} with the number of nodes in the network. In \cite{Hong-ITW}, we improved this result by using the principle of QMF (or NNC) and by optimizing the quantization levels. We showed that the gap from the capacity scales {\it logarithmically} with the number of nodes. The same rate-scaling was also attained in \cite{Kolte} by choosing quantization levels at a resolution decreasing with the number of relay stages (in short, stage-depth quantization).
Furthermore, the optimized QMF has a lower decoding complexity because it deploys successive decoding (SD) instead of joint decoding (JD) used in \cite{Avestimehr,Lim,Hou,Kolte}.

However, in network consisting of many relays, constraining all relays to perform the same scheme might not be optimal since they observe signals of different strengths. Relays in favorable positions can perform DF thereby eliminating the noise otherwise partially propagated via quantization based schemes \cite{Hou, Ivana}. On the other hand, decoding requirement at a relay can severely limit the transmission rate if the reception link is weak. Motivated by this,  we presented in \cite{Hong-ITW} a {\em mixed} strategy using both DF and QMF (with optimal quantization) for Gaussian multihop virtual full-duplex channels. We considered  a special case of this scheme restricted to a {\it symmetric} relaying configuration in which all relays on one transmission path perform DF and others in the second path perform QMF with rate-splitting.


In this paper, we generalize our previous work of \cite{Hong-ITW} in three ways: 1) we consider a general discrete memoryless channel beyond a Gaussian channel; 2) we analyze an arbitrary relaying configuration in which each relay can perform either DF or QMF to optimize the overall rate performance; 3) each relay (either DF or QMF) can employ rate-splitting that enables interference cancellation at DF relays thereby increasing the achievable rate. We derive an achievable rate of the proposed scheme. Via numerical evaluation, we show that the proposed mixed scheme outperforms the optimized QMF in \cite{Hong-ITW} as well as the QMF with noise-level \cite{Avestimehr} and stage-depth quantization \cite{Kolte}, and that the performance gain increases with the number of hops. This result implies that using DF relays in favorable positions reduces the gap from capacity to $\log{K'}$ where $K' \leq K$ denotes the number of stages that contain a QMF relay. Our results indicate that deployment of the mixed strategy for a general multiple multicast relay network can result in performance gains.

\section{Network Model}\label{sec:model}

We consider a virtual full-duplex relay channel with $K$ relay stages illustrated in Fig.~\ref{K-Hops}.
Encoding/decoding operations are performed over time slots consisting of $n$ channel uses of a discrete memoryless channel.
{\em Successive relaying} \cite{Hong-ITW} is assumed such that, at each time slot $t$, the source transmits a new message
$\underline{\wv}_{t} \in \{1,\ldots,2^{n r_{i}}\}$ where $i=1$ for odd time slot $t$ and $i=2$ for even time slot $t$, and
the destination decodes a new message $\underline{\wv}_{t-K}$. We define two message rates $r_{1}$ and $r_{2}$ since the odd-indexed and even-indexed messages are conveyed to the destination via two disjoint paths, namely, path 1: $({\rm S}, \mbox{R}_{1,1},\ldots,\mbox{R}_{1,K},{\rm D})$ and path 2: $({\rm S},\mbox{R}_{2,1},\ldots,\mbox{R}_{2,K},{\rm D})$.
The role of relays is alternatively reversed in successive time slots (see Fig.~\ref{K-Hops}).
During $N+K$ time slots, the destination decodes the $N/2$ messages from each path. Thus, the achievable rate of the messages via path $i$ is given by $r_{i}N/2(N+K)$.
By letting $N\rightarrow\infty$, the rate $r_{i}/2$ is achievable, provided that the error probability vanishes with $n$.
As in standard relay channels (see for example \cite{Avestimehr,Lim}), we take first the limit for $n \rightarrow \infty$
and then for $N \rightarrow \infty$, and focus on the achievable rate $r_{i}$.
Throughout, we use the notation $\bar{i}$ to indicate the complement of $i$, i.e., $\bar{i}=2$ if $i=1$ and $\bar{i}=1$ if $i=2$.
The discrete memoryless channel is described by the conditional probabilities given by
$\prod_{k=1}^{\lceil K/2 \rceil} p(y_{\bar{i},2k-1}|x_{i,2k-1},x_{\bar{i},2k-2})$ $\prod_{k=1}^{\lfloor K/2 \rfloor} p(y_{i,2k}|x_{i,2k-1},x_{\bar{i},2k})$ $p(y_{{\rm D}}|x_{i,K})$,
where $i=1$ for odd $t$ and $i=2$ for even $t$, and where $x_{i,k}$ and $y_{i,k}$ denote the respective output and input at relay $\mbox{R}_{i,k}$, and $x_{1,0}$ and $x_{2,0}$ denote the source outputs.

\section{Main Results}\label{sec:main}

We present a {\em mixed} scheme in which each relay performs either QMF or DF depending on channel coefficients. Each DF relay decodes its incoming message which can be either a source message or a quantization index sent by a QMF relay. The destination explicitly decodes relays’ messages as well as the source message and hence it can use these messages as a side-information in the next time slot. Furthermore, each relay can incorporate rate-splitting into its encoding scheme (QMF or DF) to reduce interference it creates to another relay. To be specific, relay $\mbox{R}_{i,k}$ uses a rate-splitting if $\mbox{R}_{\bar{i},k}$ performs DF. This enables DF relays to partially eliminate the inter-relay interference. When $\mbox{R}_{\bar{i},k}$ performs QMF, the rate-splitting is not used because a QMF relay does not need to decode any message (unlike a DF relay) and because the destination decodes a message with full-knowledge of the interference. Detailed description of the encoding/decoding scheme is given in Section~\ref{Proof:Khops}.

In order to state the achievable rate, we next introduce the following notation. Let $\Vc_{i}=\{k_{i,1},\ldots,k_{i,|\Vc_{i}|}\} \subseteq \{1,\ldots,K\}$ denote the index subset containing the indices of QMF relays in the path $i$, where $k_{i,1}<k_{i,2}<\cdots<k_{i,|\Vc_{i}|}$.
For a given $\Vc_{i}$, let $\Ic_{i,\ell} = \{k_{i,\ell},\ldots,k_{i,\ell+1}-1\}$ for $\ell=0,\ldots,|\Vc_{i}|$ with $k_{i,0} = 0$ and $k_{i,|\Vc_{i}|+1} = K+1$. Notice that $\Ic_{i,\ell}$ includes all DF relays that transmit message sent by QMF relay $\mbox{R}_{i,k_{i,\ell}}$. Define a mapping: $g_{i}(k) = k_{i,\ell}$ if $k \in \Ic_{i,\ell}$, $\ell=0,\ldots,|\Vc_{i}|$. Notice that $\{\Ic_{i,\ell}\}_{\ell=0}^{|\Vc_{i}|}$ forms a partition of $\{1,\ldots,K\}$.
\begin{definition}\label{def:I} According to the mode of a receiving relay, we define:
\begin{align*}
I_{i,k} &\eqdef \left\{
                 \begin{array}{ll}
                   I(X_{i,k};Y_{i,k+1}|U_{\bar{i},k+1}), & \hbox{$k+1 \in \Vc_{i}^{c}$}\\
                   I(X_{i,k};\hat{Y}_{i,k+1}|X_{\bar{i},k+1}), & \hbox{$k+1 \in \Vc_{i}$}
                 \end{array}
               \right.\\
I_{i,k1} &\eqdef \left\{
                 \begin{array}{ll}
                   I(X_{i,k};Y_{i,k+1}|U_{\bar{i},k+1},U_{i,k}), & \hbox{$k+1 \in \Vc_{i}^{c}$} \\
                   I(X_{i,k};\hat{Y}_{i,k+1}|X_{\bar{i},k+1},U_{i,k}), & \hbox{$k+1 \in \Vc_{i}$}
                 \end{array},
               \right.
\end{align*} where $Y_{i,K+1}=Y_{{\rm D}}$, $X_{i,K+1}=\phi$, and $U_{i,k}$ denotes an auxiliary random variable to be used for superposition coding.
\end{definition}

By letting $r_{i,k}$ denote the rate of $\mbox{R}_{i,k}$, we have:

\begin{theorem}\label{thm:khop-SD} For a $(K+1)$-hop virtual full-duplex relay channel, the achievable rate-region of the mixed strategy with SD is the set
of all rate pairs $(r_{1}/2,r_{2}/2)$ that satisfy:
\begin{equation*}
r_{i} \leq \min\left\{ \min_{k \in \Ic_{i,0}}I_{i,k},\min_{k \in \Ic_{i,0} \cap \Vc_{\bar{i}}^c}I(U_{i,k};Y_{\bar{i},k}) + I_{i,k1} \right\},
\end{equation*}
and for $k \in \Vc_{i}$ with $g_{i}(k) = k_{i,\ell}$,
\begin{align*}
&I(\hat{Y}_{i,k};Y_{i,k}|X_{\bar{i},k})\\
&= \min\left\{\min_{k'\in \Ic_{i,\ell}} I_{i,k'},\min_{k' \in \Ic_{i,\ell} \cap \Vc_{\bar{i}}^{c} }I(U_{i,k'};Y_{\bar{i},k'}) + I_{i,k'1}\right\},
\end{align*} for any index subset $\Vc_{i}\subseteq \{1,\ldots,K\}, i=1,2,$ and any joint distributions that factors as $\prod_{i=1}^{2}p(x_{i,0})$ $\prod_{k \in \Vc_{\bar{i}}}p(x_{i,k})$$\prod_{k \in \Vc_{\bar{i}}^{c}}p(u_{i,k})p(x_{i,k}|u_{i,k})$$\prod_{k \in \Vc_{i}}p(\hat{y}_{i,k}|y_{i,k})$.
\end{theorem}
\begin{IEEEproof}
See Section~\ref{Proof:Khops}.
\end{IEEEproof}

\begin{theorem}\label{cor1} For a $(K+1)$-hop virtual full-duplex relay channel, the achievable rate region of the mixed strategy with JD (at DF-only stages) is the set of all rate pairs $(r_{1}/2,r_{2}/2)$ to satisfy:
\begin{equation*}
r_{i} \leq  \min \{I_{i,k}: k\in \Ic_{i,0}\},
\end{equation*}and for $k \in \Vc_{1}^{c} \cap \Vc_{2}^{c}$,
\begin{align*}
&r_{1,g_{1}(k)} + r_{2,g_{2}(k)} \leq \min\{I(U_{1,k},X_{2,k-1};Y_{2,k})+ I_{1,k1},\\
&\qquad\qquad\qquad\qquad I(U_{2,k},X_{1,k-1};Y_{1,k}) + I_{2,k1}\},
\end{align*}
 and for $k \in \Vc_{i}$ with $g_{i}(k) = k_{i,\ell}$,
\begin{align*}
&I(\hat{Y}_{i,k};Y_{i,k}|X_{\bar{i},k}) = r_{i,k}\\
&r_{i,k} \leq \min\{I_{i,k'}: k' \in \Ic_{i,\ell}\}\\
&r_{i,k} \leq I(U_{i,k};Y_{\bar{i},k}) + I_{i,k1}, k \in \Vc_{\bar{i}}^{c},
\end{align*} for any subset $\Vc_{i} \subseteq \{1,\ldots,K\}$ and any joint distribution given in Theorem 1, where $r_{i,0} = r_{i}$.
\end{theorem}
\begin{IEEEproof} The proof is omitted due to the lack of space.
\end{IEEEproof}

\begin{remark}\label{remark1} Recall that JD is a key component in the Han-Kobayashi coding scheme \cite{Han} for two-user interference channel, that achieves a higher rate than SD. In other words, JD attains a higher rate than SD when both messages of two transmitters should be decoded at two receivers, i.e., the corresponding rates should be chosen in the intersection of two MAC regions. In our network, this case occurs when both relays at the same stage perform DF. Thus, we applied JD for such stages to obtain an improved achievable rate in Corollary~\ref{cor1}.
\end{remark}

\subsection{Proof of Theorem~\ref{thm:khop-SD}}\label{Proof:Khops}

\begin{figure}
\centerline{\includegraphics[width=8cm]{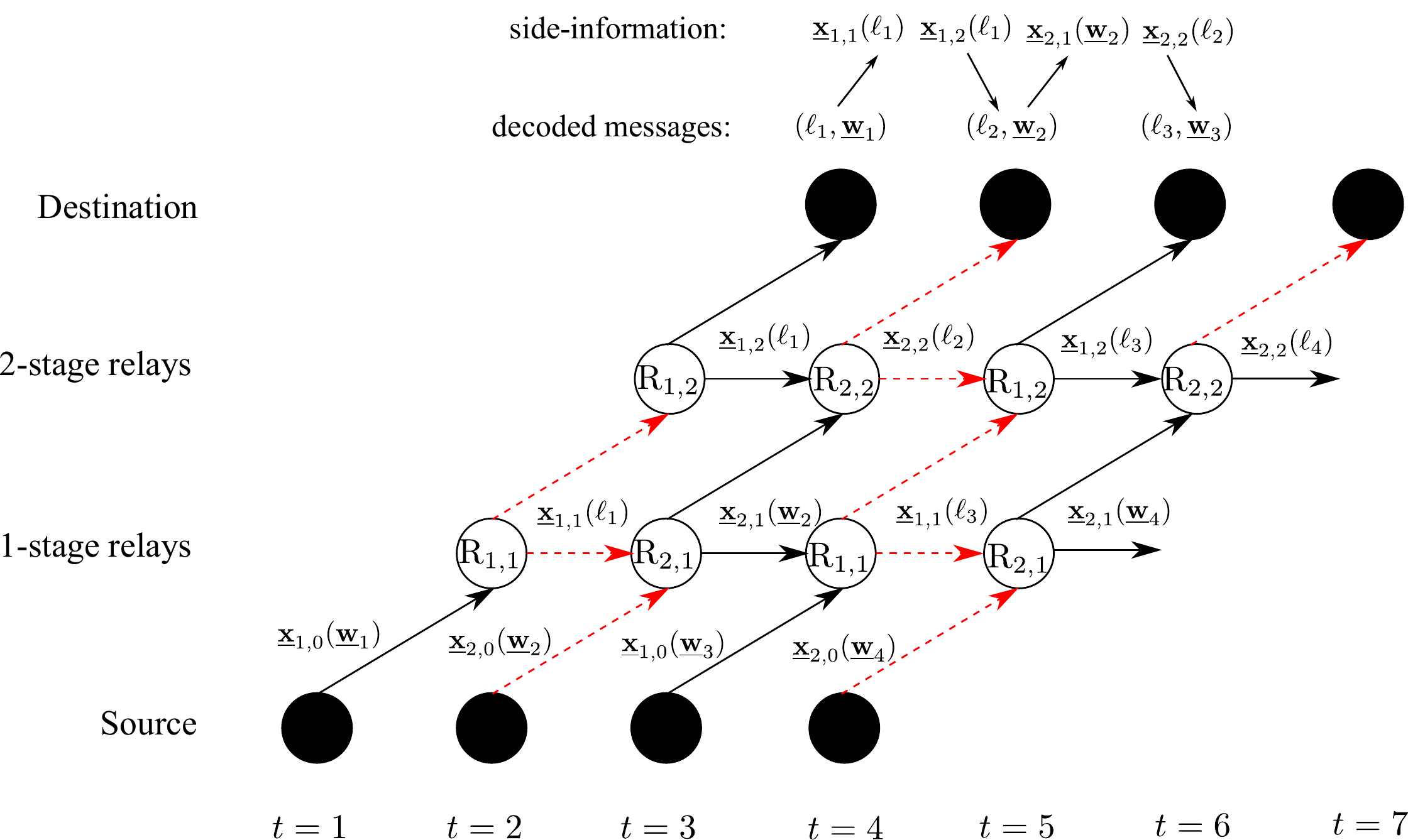}}
\caption{Time expanded 3-hop network where relays $\mbox{R}_{1,2}$ and $\mbox{R}_{2,1}$ perform DF and others perform QMF.
$l_i$ denotes the quantization index obtained at a QMF relay in time slot $i$.}
\label{TimeEx}
\end{figure}

\begin{figure}
\centerline{\includegraphics[width=6cm]{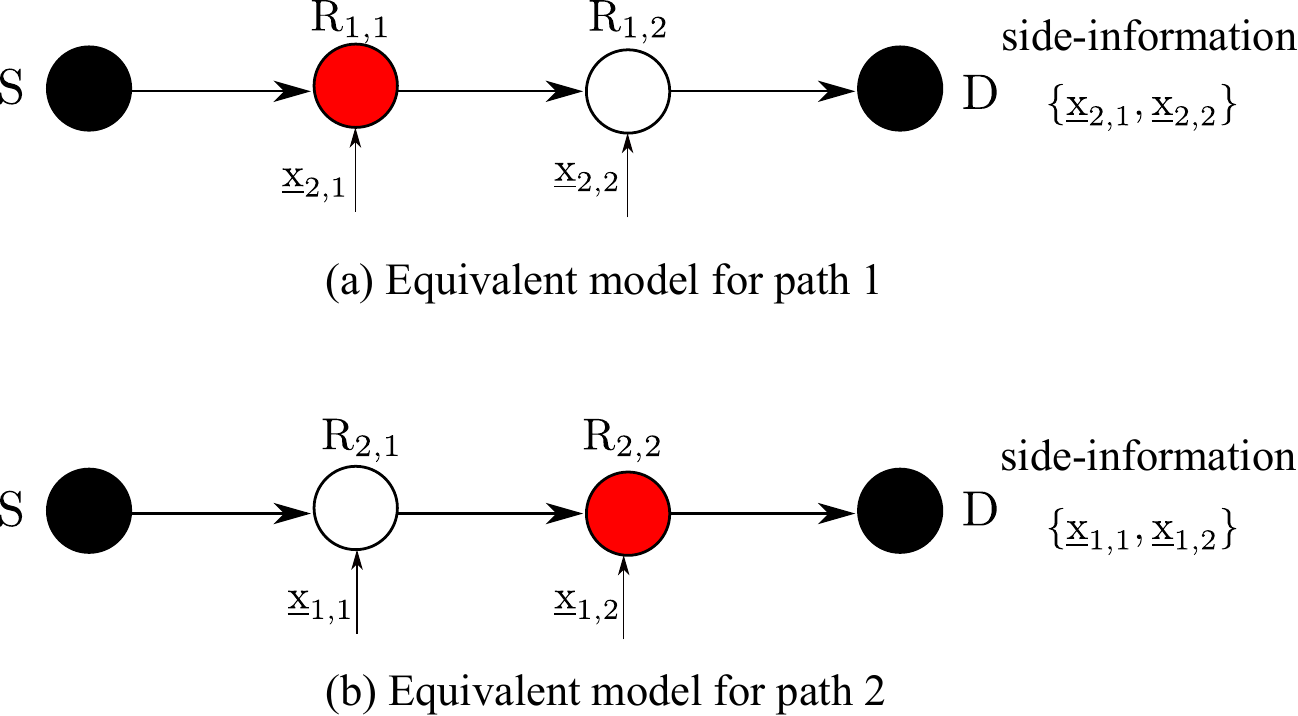}}
\caption{Equivalent model for 3-hop network where red circles perform QMF and white circles perform DF.}
\label{Equiv}
\end{figure}

Fig.~\ref{TimeEx} shows a {\em time-expanded} graph of a 3-hop virtual full-duplex relay channel in which relays $\mbox{R}_{1,2}$ and $\mbox{R}_{2,1}$ perform DF and  $\mbox{R}_{1,1}$ and $\mbox{R}_{2,2}$ perform QMF. As pointed out earlier, in the proposed scheme, the destination explicitly decodes relays' messages and hence, it can use these messages as side-information in the next time slot thereby completely knowing the inter-relay interference (see Fig.~\ref{TimeEx}). From this, we can produce a simplified network model shown in Fig.~\ref{Equiv}, which can be straightforwardly extended to a $(K+1)$-hop network with an arbitrary relay configuration. This network model will be used for the proof.

Fix relay modes $\Vc_{i}$, $i=1,2$. For given $\Vc_{i}$, fix input distributions as defined in Theorem~\ref{thm:khop-SD}.

\textbf{Codebook generation:} Randomly and independently generate $2^{n r_{i}}$ codewords $\underline{\xv}_{i,0}(w_{i})$ of length $n$ indexed by $w_{i} \in \{1,\ldots,2^{n r_{i}}\}$ with i.i.d. components $\sim p(x_{i,0})$. For $k \in \Vc_{\bar{i}}$ (i.e., QMF interfered relay), randomly and independently generate $2^{n r_{i,k}}$ codewords $\underline{\xv}_{i,k}(\ell_{i,k})$ of length $n$ indexed by $\ell_{i,k} \in \{1,\ldots,2^{n r_{i,k}}\}$ with i.i.d. components $\sim p(x_{i,k})$. For $k \in \Vc_{\bar{i}}^{c}$ (i.e., DF interfered relay), split message $\ell_{i,k}$ into independent ``common" message $\ell_{i,k}^{c}$ at rate $r_{i,k0}$ and ``private" message $\ell_{i,k}^{p}$ at rate $r_{i,k1}$ with $r_{i,k}=r_{i,k1}+r_{i,k0}$. Randomly and independently generate $2^{n r_{i,k0}}$ codewords $\underline{\uv}_{i,k}(\ell_{i,k}^{c})$ of length $n$ indexed by $\ell_{i,k}^{c} \in \{1,\ldots,2^{n r_{i,k0}}\}$ with i.i.d. components $\sim p(u_{i,k})$. For each $\ell_{i,k}^{c}$, randomly and conditionally independently generate $2^{n r_{i,k1}}$ codewords $\underline{\xv}_{i,k}(\ell_{i,k}^{c},\ell_{i,k}^{p})$ of length $n$ indexed by $\ell_{i,k}^{p} \in \{1,\ldots,2^{n r_{i,k1}}\}$ with i.i.d. components $\sim p(x_{i,k}|u_{i,k})$. Randomly and independently generate $2^{n\hat{r}_{i,k}}$ codewords $\hat{\underline{\yv}}_{i,k}(\mu)$ of length $n$ indexed by $\mu \in \{1,\ldots,2^{n\hat{r}_{i,k}}\}$ with i.i.d. components $\sim p(\hat{y}_{i,k})$. The quantization codewords are randomly and independently assigned with uniform probability to $2^{n r_{i,k}}$ bins. Denote the $\ell_{i,k}$-th bin by $\Bc(\ell_{i,k})$ with $\ell_{i,k} \in \{1,\ldots,2^{n r_{i,k}}\}$. Here, Wyner-Ziv quantization is assumed, such that the quantization distortion level is chosen by imposing $I(\hat{Y}_{i,k};Y_{i,k}|X_{\bar{i},k}) = r_{i,k}$.
This enables the destination to find an unique quantization sequence $\hat{\yv}_{i,k}$ from the bin index $\ell_{i,k}$ and the side-information $\underline{\xv}_{i,k}$.

\textbf{Encoding:} Source transmits a message $w_{i}$ by sending the codeword $\underline{\xv}_{i,0}(w_{i})$ where $i$ is either $1$ or $2$ depending on time slot. For QMF relay with $k \in \Vc_{i}$, $\mbox{R}_{i,k}$ observes $\underline{\yv}_{i,k}$ and finds $\mu$ such that $(\underline{\yv}_{i,k},\hat{\underline{\yv}}_{i,k}(\mu)) \in \Tc_{\epsilon}^{(n)}(Y_{i,k},\hat{Y}_{i,k})$. If no quantization codeword satisfies the joint typicality condition, the relay chooses $\mu=1$. Then, it finds the bin index $\ell_{i,k}$ such that $\hat{\underline{\yv}}_{i,k}(\mu) \in \Bc(\ell_{i,k})$. To send the message $\ell_{i,k}=(\ell_{i,k}^{c},\ell_{i,k}^{p})$, it transmits the downstream codeword $\underline{\xv}_{i,k}(\ell_{i,k}^{c}, \ell_{i,k}^{p})$ using superposition coding. If rate-splitting is not used, then the codeword $\underline{\xv}_{i,k}(\ell_{i,k})$ is sent. For DF relay with $k \in \Vc_{i}^{c}$, $\mbox{R}_{i,k}$ decodes the incoming message $\hat{\ell}_{i,k-1}$. To send the message $\hat{\ell}_{i,k-1}=(\hat{\ell}_{i,k-1}^{c},\hat{\ell}_{i,k-1}^{p})$, it transmits the downstream codeword $\underline{\xv}_{i,k}(\hat{\ell}_{i,k-1}^{c}, \hat{\ell}_{i,k-1}^{p})$ using superposition coding. If the rate-splitting is not used, then the codeword $\underline{\xv}_{i,k}(\hat{\ell}_{i,k-1})$ is sent.

\begin{figure}
\centerline{\includegraphics[width=8cm, height=2.6cm]{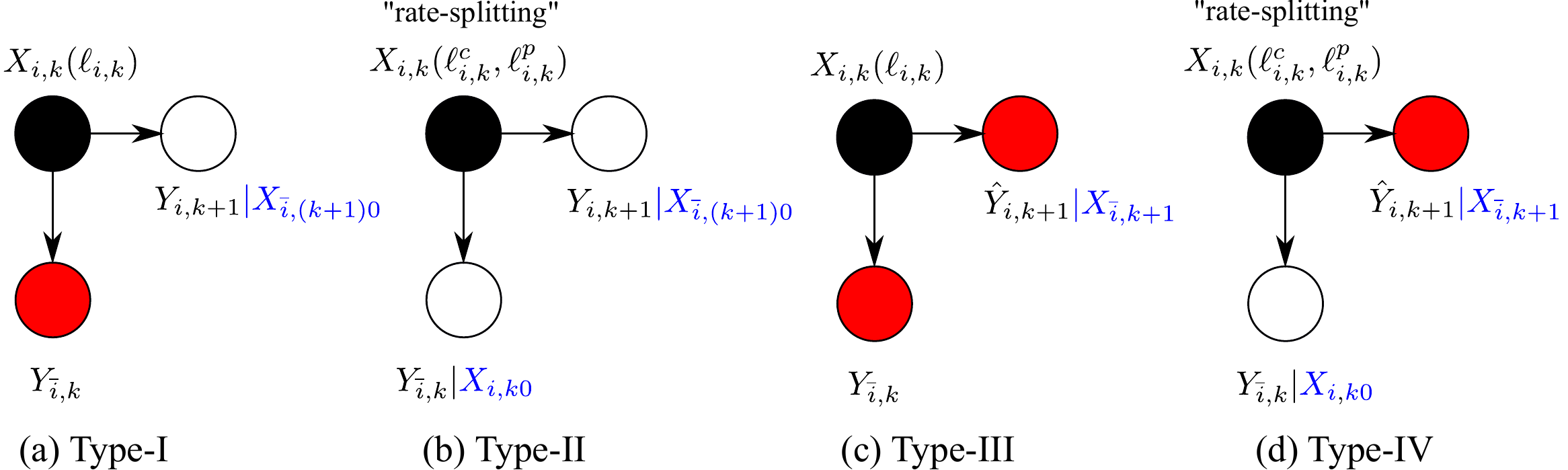}}
\caption{Four scenarios determined by the next-hop and interfered relays. Red circles perform QMF, white circles perform DF, and black circles perform either QMF or DF.}
\label{khop-type}
\end{figure}

\textbf{Decoding:} We first observe that rate $r_{i,k}$ at relay $\mbox{R}_{i,k}$ is determined according to the modes of its neighboring relays, i.e., the next-hop relay $\mbox{R}_{i,k+1}$ and the interfered relay $\mbox{R}_{\bar{i},k}$. When $\mbox{R}_{i,k+1}$ performs QMF, the destination decodes $\hat{\ell}_{i,k}$ from a quantized observation $\hat{\underline{\yv}}_{i,k+1}$ with the full-knowledge of the interference $\underline{\xv}_{\bar{i},k+1}$. In the other case, $\mbox{R}_{i,k+1}$ decodes the $\hat{\ell}_{i,k}$ from its observation $\underline{\yv}_{i,k+1}$ with the partial-knowledge of interference $\underline{\uv}_{\bar{i},k+1}$. In addition, the mode of $\mbox{R}_{\bar{i},k}$ determines the use of rate-splitting at $\mbox{R}_{i,k}$, yielding an additional rate-constraint of $r_{i,k0}$ since the common message $\ell_{i,k}^{c}$ should be decoded at $\mbox{R}_{\bar{i},k}$. Namely, the modes of the neighboring relays determines the types of observations (unquantized vs quantized), side-information (full-knowledge vs partial-knowledge), and the use of rate-splitting. Based on this, we can define the four scenarios given in Fig.~\ref{khop-type}. The corresponding rate-constraints of $r_{i,k}$ are derived as follows.

{\it Type-I:} Since the interfered relay $\mbox{R}_{\bar{i},k}$ performs QMF, $\mbox{R}_{i,k}$ does not use rate-splitting as explained before. Due to the rate-splitting at $\mbox{R}_{\bar{i},k+1}$, the next-hop relay $\mbox{R}_{i,k+1}$ can reliably decode the $\ell_{i,k}$  with the partial-knowledge of interference as $\underline{\uv}_{\bar{i},k+1}$, yielding:
\begin{equation}
r_{i,k} \leq I(X_{i,k};Y_{i,k+1}|U_{{\bar i},k+1})\label{eq-type:const1}.
\end{equation}

{\it Type-II:} In this case, $\mbox{R}_{i,k}$ uses rate-splitting so that the interfered relay $\mbox{R}_{\bar{i},k}$ can partially eliminate the interference. Thus, the common message $\ell_{i,k0}$ should be decoded at $\mbox{R}_{\bar{i},k}$, yielding
\begin{equation}
r_{i,k 0} \leq I(U_{i,k};Y_{\bar{i},k}).\label{eq:SD}
\end{equation}
The next-hop relay $\mbox{R}_{i,k+1}$ can reliably decode the $\ell_{i,k}=(\ell_{i,k}^{c},\ell_{i,k}^{p})$ if
\begin{align*}
&r_{i,k 0} \leq I(U_{i,k};Y_{i,k+1}|U_{\bar{i},k+1})\\
&r_{i,k 1} \leq I(X_{i,k};Y_{i,k+1}|U_{i,k},U_{\bar{i},k+1}).
\end{align*} From the above, we get:
\begin{align}
r_{i,k} &\leq \min\{I(X_{i,k};Y_{i,k+1}|U_{\bar{i},k+1}),\label{eq-type:const2}\\
&I(U_{i,k};Y_{\bar{i},k})+I(X_{i,k};Y_{i,k+1}|U_{i,k},U_{\bar{i},k+1})\}.\nonumber
\end{align}

{\it Type-III:} The destination can reliably decode the $\ell_{i,k}$ from a quantized observation $\hat{\yv}_{i,k+1}$ using the side-information $\xv_{{\bar i},k+1}$ if
\begin{align}
&r_{i,k} \leq I(X_{i,k};\hat{Y}_{i,k+1}|X_{\bar{i},k+1})\label{eq-type:const3}.
\end{align}

{\it Type-IV:} With the same argument in Type-II, $\mbox{R}_{i,k}$ uses rate-splitting and hence, the common message $\ell_{i,k}^{c}$ should be decoded at $\mbox{R}_{\bar{i},k}$, yielding
\begin{equation}
r_{i,k 0} \leq I(U_{i,k};Y_{\bar{i},k}).\label{eq:SD1}
\end{equation} The destination can reliably decode the $\ell_{i,k}=(\ell_{i,k}^{c},\ell_{i,k}^{p})$ if
\begin{align*}
&r_{i,k 0} \leq I(U_{i,k};\hat{Y}_{i,k+1}|X_{\bar{i},(k+1)})\\
&r_{i,k 1} \leq I(X_{i,k};\hat{Y}_{i,k+1}|U_{i,k},X_{\bar{i},(k+1)}).
\end{align*} From the above, we obtain:
\begin{align}
&r_{i,k} \leq\min\{I(X_{i,k};\hat{Y}_{i,k+1}|X_{\bar{i},(k+1)}),\label{eq-type:const5}\\
&\qquad\qquad I(U_{i,k};Y_{\bar{i},k})+I(X_{i,k};\hat{Y}_{i,k+1}|U_{i,k},X_{\bar{i},(k+1)})\}\nonumber
\end{align}

We are now ready to derive an achievable rate of the mixed scheme. From Fig.~\ref{khop-type}, we can classify the types of each relay $\mbox{R}_{i,k}$.
Using (\ref{eq-type:const1})-(\ref{eq-type:const5}) and from Definition~\ref{def:I}, we have:
\begin{align*}
&r_{i,k} \leq \left\{
            \begin{array}{ll}
              I_{i,k}, & \mbox{{\it Types I and III}}\\
              \min\{I_{i,k}, I(U_{i,k};Y_{\bar{i},k})+I_{i,k1}\}, & \mbox{{\it Types II and IV}}
            \end{array}
          \right.
\end{align*} which can be represented as
\begin{align}
r_{i,k} &\leq I_{i,k}\label{eq:constf1} \\
r_{i,k} &\leq I(U_{i,k};Y_{\bar{i},k})+I_{i,k1}, k \in \Vc_{\bar{i}}^{c}.\label{eq:constf2}
\end{align} Then, we have:
\begin{itemize}
\item The source and DF relays $\mbox{R}_{i,k}$ for $k \in \Ic_{i,0}$ send a source message. This message can be reliably decoded at those DF relays and the destination if
\begin{equation}
r_{i} \leq \min\{r_{i,k}:k \in \Ic_{i,0} \}.\label{eq:ratef1}
\end{equation}
\item Let $g_{i}(k) = k_{i,\ell}$. The relay's message $\ell_{i,k}$ can be decoded at DF relays $\mbox{R}_{i,k'}$ for $k' \in \Ic_{i,\ell}$ and the destination if
\begin{equation}
r_{i,k} \leq \min\{r_{i,k'}:k' \in \Ic_{i,\ell}\}.\label{eq:ratef2}
\end{equation}
\item Due to the use of Wyner-Ziv quantization, we have:
\begin{equation*}
I(\hat{Y}_{i,k};Y_{i,k}|X_{\bar{i},k})=r_{i,k}, k \in \Vc_{i}.
\end{equation*}
\end{itemize} Substituting (\ref{eq:constf1}) and (\ref{eq:constf2}) into (\ref{eq:ratef1}) and (\ref{eq:ratef2})  completes the proof.

\begin{figure}[t]
\centerline{\includegraphics[width=8cm]{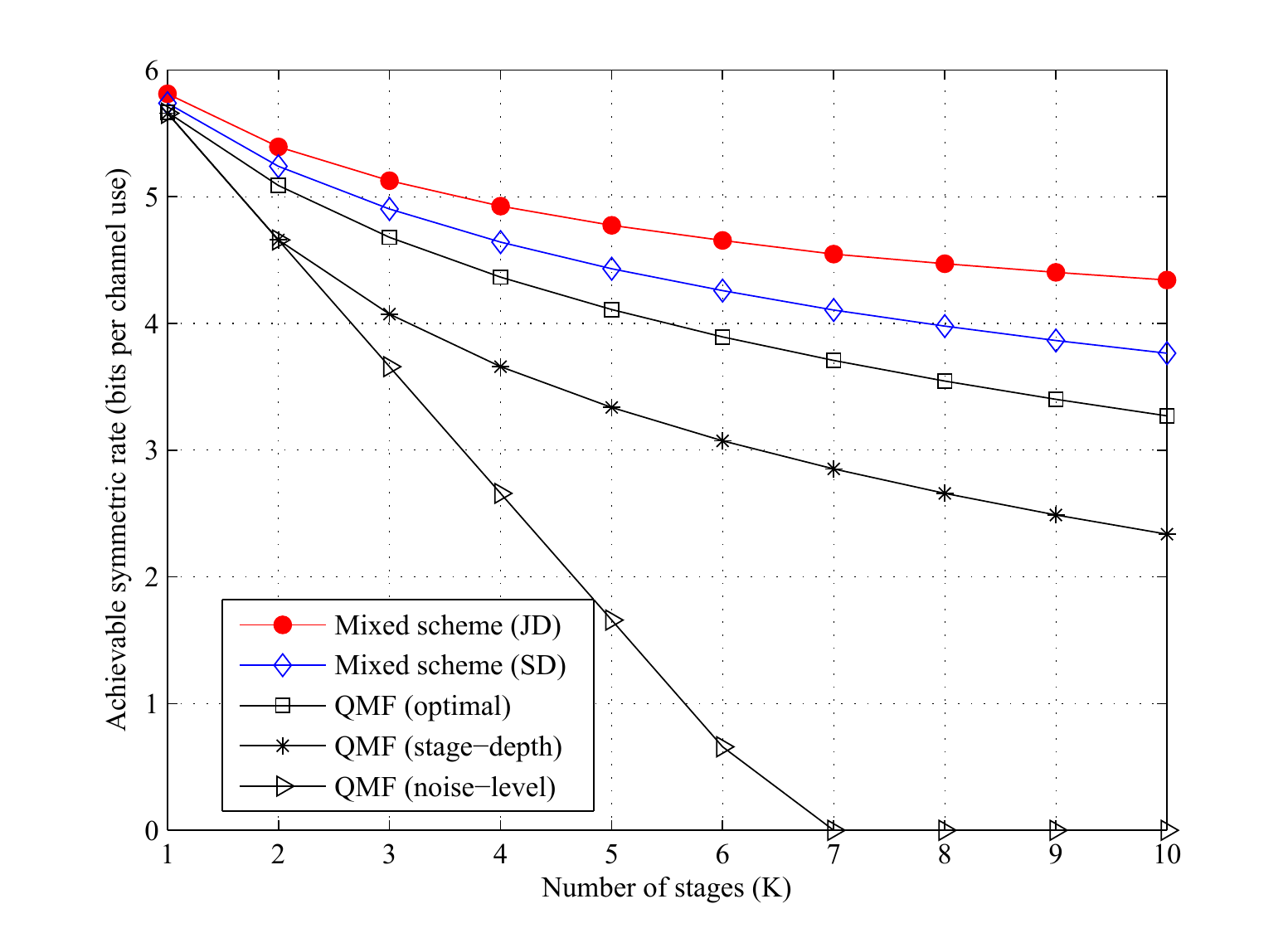}}
\caption{$\SNR=20$dB and $\INR_{k} =\SNR^{\alpha_{k}}$ where $\alpha_{k} \sim \mbox{Unif}[1,2]$.}
\label{simulation1}
\end{figure}

\section{Achievable rates for Gaussian Channels}\label{sec:GC}

In order to evaluate the performance of the proposed scheme, we consider a Gaussian channel where both paths experience the same channel gains, i.e., $\SNR_{k}$ denotes the direct channel gain from $\mbox{R}_{i,k-1}$ to $\mbox{R}_{i,k}$ and $\INR_{k}$ denotes the interference channel gain from  $\mbox{R}_{\bar{i},k}$ to $\mbox{R}_{i,k}$. Due to the symmetric structure of each stage, we naturally assume that each stage uses the same relaying scheme, i.e., $\Vc_{1}=\Vc_{2}\eqdef \Vc=\{k_{1},\ldots,k_{|\Vc|}\}$. From Definition~\ref{def:I}, we  obtain:
\begin{align*}
&I_{k} = \log\left(1+\frac{\SNR_{k+1}}{1+(1-\theta_{k+1})\INR_{k+1}+\hat{\sigma}_{k+1}^2}\right)\\
&I_{k1} = \log\left(1+\frac{(1-\theta_{k})\SNR_{k+1}}{1+(1-\theta_{k+1})\INR_{k+1}+\hat{\sigma}_{k+1}^2}\right),
\end{align*} where $\theta_{k+1}=1$ if $k+1 \in \Vc$ and $\hat{\sigma}_{k+1}^2=0$ if $k+1 \in \Vc^{c}$.  In this section, we focus on the symmetric achievable rate of $r$ with $r=r_{1}=r_{2}$. We obtain:
\begin{corollary}\label{cor2} For a $(K+1)$-hop Gaussian virtual full-duplex relay channel, the achievable symmetric rate of the mixed strategy with SD (or JD) is given by
\begin{align*}
r =& \min\{\min\{I_{k}: k \in \Ic_{0}\}, \min\{I'_{k}: k \in \Ic_{0}\setminus\{0\}\}\}\\
r_{k_{\ell}} =& \min\{\min\{I_{k}: k \in \Ic_{\ell}\}, \min\{I'_{k}: k \in \Ic_{\ell}\setminus\{k_{\ell}\}\}\}\\
\hat{\sigma}_{k_{\ell}}^2 =& (1+\SNR_{k_{\ell}})/ (2^{r_{k_{\ell}}}-1), \ell=1,\ldots,|\Vc|,
\end{align*}for any subset $\Vc \subseteq \{1,\ldots,K\}$ and any $\theta_{k} \in [0,1]$ with $\theta_{k}=1$ for $k \in \Vc$, where
\begin{align*}
I'_{k} = \left\{
           \begin{array}{ll}
             \log\left(1+\frac{\theta_{k+1}\INR_{k+1}}{1+\SNR_{k+1}}\right) + I_{k1}, & \hbox{SD} \\
             \frac{1}{2}\log\left(1+\frac{\SNR_{k+1}+\theta_{k+1}\INR_{k+1}}{1+(1-\theta_{k+1})\INR_{k+1}}\right) + \frac{1}{2}I_{k1}, & \hbox{JD.}
           \end{array}
         \right.
\end{align*}
\end{corollary}
\begin{IEEEproof} The proof follows the proof of Theorem~\ref{cor1} by choosing Gaussian input distributions with the conventional power-splitting approach and by setting $r_{1,k}=r_{2,k}$ for $k=1,\ldots,K$. The detailed proof is omitted.
\end{IEEEproof}

\begin{figure}[t]
\centerline{\includegraphics[width=8cm]{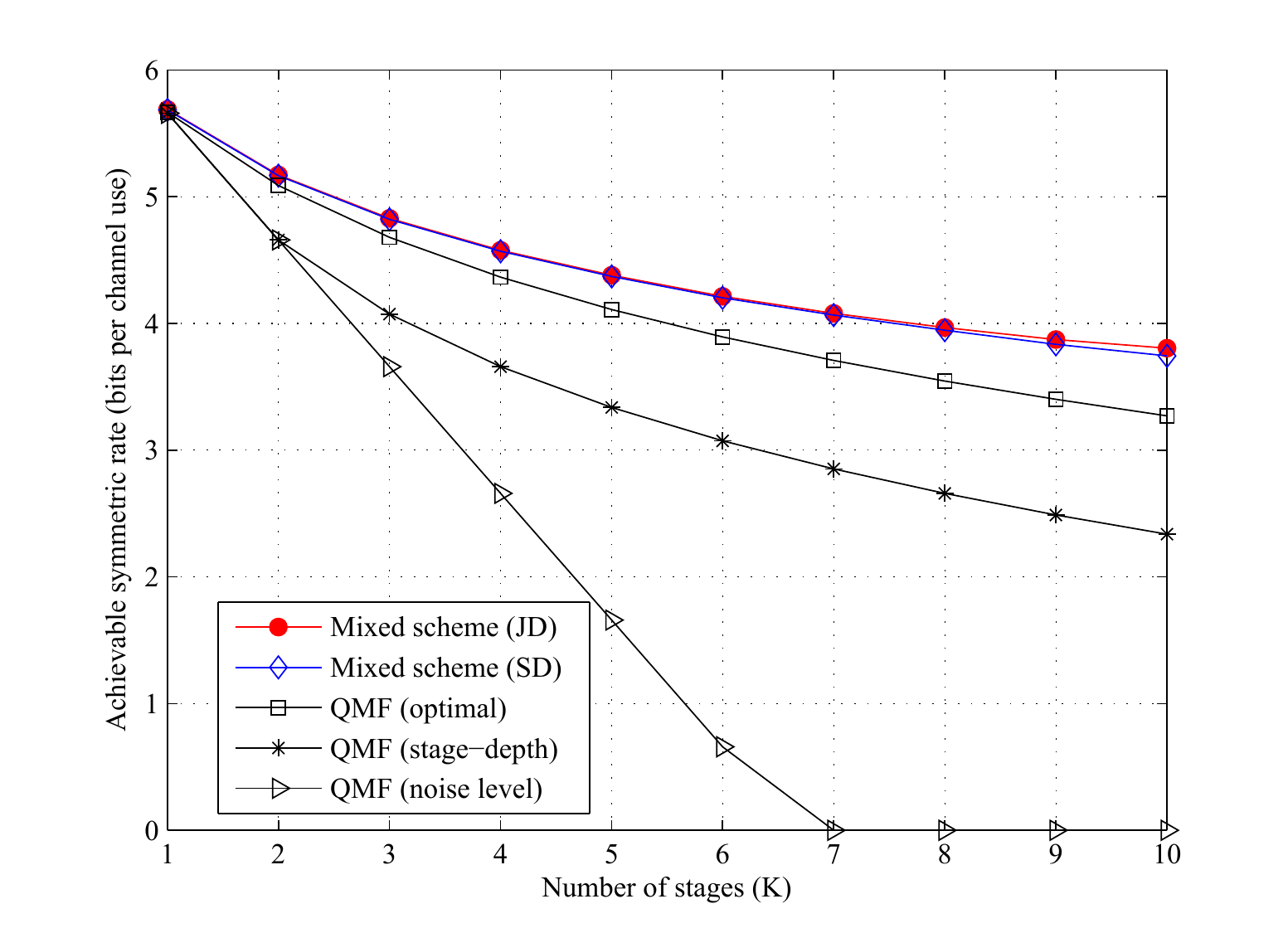}}
\caption{$\SNR=20$dB and $\INR_{k} =\SNR^{\alpha_{k}}$ where $\alpha_{k} \sim \mbox{Unif}[0,1]$.}
\label{simulation2}
\end{figure}

In Figs.~\ref{simulation1} and~\ref{simulation2}, we numerically evaluate the achievable symmetric rate of the proposed scheme for different values of $K$. Here, we performed an exhaustive search (i.e., considered $2^{K}$ possible configurations) to find the best $\Vc$ and power-splitting parameters $\theta_{k}$. For comparison, we consider the performance of QMF with various quantization levels as noise-level \cite{Avestimehr}, stage-depth \cite{Kolte}, and optimal quantization \cite{Hong-ITW}. Our results show that the proposed mixed scheme outperforms the QMF schemes achieving a larger gap as $K$ grows. Furthermore, we confirmed the argument in Remark~\ref{remark1} by showing, in Fig.~\ref{simulation1}, that JD significantly improves the performance compared with SD.
By comparing Figs.~\ref{simulation1} and~\ref{simulation2}, we observe that this gain is larger in strong inter-relay interference since in weak interference, we treat interference as noise in both cases.

\section{Conclusion}

In this work, we proposed a mixed scheme for multihop ``virtual" full-duplex relay channels. We showed that our scheme outperforms the optimized QMF and, furthermore, that this improvement increases with the number of hops. This implies that using DF relays in favorable positions can reduce the gap from the capacity to $\log{K'}$ where $K' \leq K$ denotes the number of stages containing a QMF relay. Based on the obtained results, we expect that the proposed mixed scheme can bring performance gains in a general multiple multicast relay network. This is a subject of our future work.
%

\end{document}